\documentclass[12pt,a4paper]{article}
\pdfoutput=1
\usepackage{jcappub}
\usepackage{amsmath}
\usepackage{latexsym}
\usepackage{amsfonts}
\usepackage{graphicx}
\usepackage{mathrsfs}
\usepackage{epstopdf}
\usepackage{orcidlink}

\title{Observational constraints on inflationary models with non-minimally derivative coupling by ACT}

\author{Qing Gao$^{a}$ \orcidlink{0000-0003-3797-4370},}
\author{Yanjiang Qian$^{a}$,}
\author{Yungui Gong$^{b}$ \orcidlink{0000-0001-5065-2259}}
\author{and Zhu Yi$^{c,1}$ \orcidlink{0000-0001-7770-9542}\note{Corresponding author}}

\affiliation{$^{a}$ School of Physical Science and Technology, Southwest University, 2 Tiansheng Rd, Chongqing 400715, China}
\affiliation{$^{b}$ Institute of Fundamental Physics and Quantum Technology, Department of Physics, School of Physical Science and Technology, Ningbo University, 818 Fenghua Rd, Ningbo, Zhejiang 315211, China}
\affiliation{$^{c}$ Faculty of Arts and Sciences, Beijing Normal University, 18 Jinfeng Rd,  Zhuhai 519087, China}

\emailAdd{gaoqing1024@swu.edu.cn}
\emailAdd{gongyungui@nbu.edu.cn}
\emailAdd{yz@bnu.edu.cn}

\abstract{The most recent data release from the Atacama Cosmology Telescope (ACT) reveals a larger value of the scalar spectral tilt $n_s$, ruling out a broad class of inflationary attractors.
In this paper, we consider inflationary models including the power law potential, the hilltop model, the polynomial $\alpha$-attractor and exponential $\alpha-$attractor, with non-minimally derivative coupling in the high friction limit,
and show how the models can fit ACT data.
We also derive constraints on the model parameters using the latest ACT data.
}

\begin{document}
\maketitle

\section{Introduction}
The latest data release from the Atacama Cosmology Telescope (ACT) \cite{Louis:2025tst,ACT:2025tim} reveals a larger value of the scalar spectral tilt $n_s$, ruling out a broad class of inflationary models including the Starobinsky model \cite{Starobinsky:1980te}, Higgs inflation \cite{Kaiser:1994vs, Bezrukov:2007ep} and $\alpha$-attractors \cite{Kallosh:2013hoa,Kallosh:2013maa},
at the $2\sigma$ confidence level.
The Planck 2018 measurements of the Cosmic Microwave Background (CMB) anisotropies reported
$n_s=0.9651\pm 0.0044$ which is consistent with the attractors $n_s=1-2/N$ \cite{Lin:2015fqa,Yi:2016jqr,Fei:2017fub,Gao:2017uja}, and an upper bound on the tensor-to-scalar ratio $r<0.101$ at $95\%$ confidence \cite{Planck:2018jri,Planck:2018vyg}.
However, a combined analysis of Planck and ACT data (hereafter P-ACT) yields a larger scalar spectral tilt, with $n_s = 0.9709 \pm 0.0038$ \cite{Louis:2025tst}.
When further including CMB lensing and Baryon Acoustic Oscillation (BAO) distance measurements from the Dark Energy Spectroscopic Instrument (DESI) \cite{DESI:2024uvr, DESI:2024mwx} (P-ACT-LB), the value increases to $n_s = 0.9743 \pm 0.0034$ \cite{Louis:2025tst},
representing a $\sim 2\sigma$ deviation from the original Planck 2018 result.
Adding B-mode polarization data from the BICEP/Keck telescopes (BK18) \cite{BICEP:2021xfz} at the South Pole to this combined dataset (P-ACT-LB-BK18) tightens the upper limit on the tensor-to-scalar ratio to $r < 0.038$ at $95\%$ confidence level \cite{ACT:2025tim}.

The results suggest that the inflationary models with higher values of $n_s$ and lower values of $r$ are favored.
In particular, the ACT data appear to rule out  monomial inflation models with convex potentials within the framework of Einstein's general relativity (GR) \cite{ACT:2025tim}.
These ACT findings have revitalized interest in alternative scenarios, including models predicting larger values of $n_s$ and those with nonminimal coupling \cite{Gialamas:2025kef,Frob:2025sfq,Dioguardi:2025vci,Brahma:2025dio,Berera:2025vsu,Aoki:2025wld,Dioguardi:2025mpp,Salvio:2025izr, Liu:2025qca, Haque:2025uis, Zharov:2025evb, Haque:2025uri, Drees:2025ngb,He:2025bli,Yogesh:2025wak, Gialamas:2025ofz,Peng:2025bws,Mondal:2025kur,Yi:2025dms,Addazi:2025qra,Maity:2025czp,Byrnes:2025kit,Kallosh:2025rni,Gao:2025onc,Katsoulas:2025mcu,Wolf:2025ecy,Odintsov:2025wai,Pallis:2025nrv}.

A widely studied class of inflationary models with nonminimal coupling involves a scalar field coupled to the Ricci scalar via a term of the form $f(\phi)R$,
which is a specific case of the more general scalar-tensor theory $F(\phi,R)$ \cite{Guo:2010jr,Jiang:2013gza,Boubekeur:2015xza}.
This type of coupling can be removed through a conformal transformation, reducing the theory to GR with a redefined scalar sector.
However, if the kinetic term of the scalar field is non-minimally coupled to the curvature tensor,
a conformal transformation cannot recast the model into a scalar-tensor theory \cite{Amendola:1993uh}.
Generally, non-minimally derivative couplings introduce higher than second-order derivatives in the field equations and extra degrees of freedom,
potentially leading to the Boulware-Deser ghost \cite{Boulware:1972zf}.
A particular choice, where the derivative coupling takes the form $G^{\mu\nu}\phi_{,\mu}\phi_{,\nu}$ which is a special case of the most general Horndeski theory \cite{Horndeski:1974wa}, avoids this issue because the resulting field equations in Horndeski theory contain no higher than second derivatives \cite{Horndeski:1974wa,Sushkov:2009hk}.
Moreover, the new Higgs inflation model, through the introduction of the non-minimally derivative coupling, satisfies the unitarity bound \cite{Germani:2010gm,Germani:2010ux}. The gravitationally enhanced friction resulting from this coupling also slows down the evolution of the scalar field \cite{Germani:2010gm,Germani:2010ux,Tsujikawa:2012mk,Yang:2015pga}.
In this paper, we show that various inflationary models with such non-minimally derivative coupling are well consistent with the latest ACT data.

The paper is organized as follows. In Section \ref{sec2},
we introduce the non-minimally derivative coupling model and present the expressions for the slow-roll parameters under the high friction condition.
We then review the scalar and tensor perturbations within this context.
In Section \ref{sec3}, we derive the expressions for the scalar spectral tilt
$n_s$ and the tensor-to-scalar ratio $r$  for various inflationary models featuring non-minimal derivative coupling.
A detailed comparison between numerical results and theoretical predictions for both the non-minimally derivative coupling scenario and the standard GR case is provided.
Additionally, we use the latest ACT data release to constrain the model parameters. Finally, our conclusions are summarized in Section \ref{sec4}.

\section{The model with non-minimally derivative coupling}
\label{sec2}

The action for the scalar field with the kinetic term non-minimally coupled to the Einstein tensor is \cite{Sushkov:2009hk,Germani:2010gm,Germani:2010ux}
\begin{equation}
\label{eq1}
S = \frac{1}{2}\int d^4 x\sqrt {-g} \left[R - g^{\mu \nu }\partial _\mu \phi {\partial _\nu }\phi  + \frac{1}{{{M^2}}}{G^{\mu \nu }}{\partial _\mu }\phi {\partial _\nu }\phi  - 2V(\phi) \right],
\end{equation}
where the Einstein tensor $G_{\mu\nu}=R_{\mu\nu}-g_{\mu\nu}R/2$, $R_{\mu\nu}$ is the Ricci tensor, $R$ is the Ricci scalar, $M$ is the coupling constant with the dimension of mass and we take  the Planck mass $M_{pl}^2 = (8\pi G)^{-1} =1$.
From the action \eqref{eq1}, we get the Friedmann equation and the
equation of motion for the scalar field $\phi$
\begin{equation}
\label{eq2}
H^2=\left(\frac{\dot a}{a}\right)^2 = \frac{1}{3}\left[\frac{\dot \phi^2}{2}\left(1  + 9\frac{H^2}{M^2} \right) + V(\phi) \right],
\end{equation}
\begin{equation}
\label{eq3}
\frac{d}{{dt}}\left[ {{a^3}\dot \phi \left( {1  + 3\frac{{{H^2}}}{{{M^2}}}} \right)} \right] =  - {a^3}\frac{{dV}}{{d\phi }}.
\end{equation}
We see that the non-minimally derivative coupling $G^{\mu\nu}\phi_{,\mu}\phi_{,\nu}/M^2$ which becomes $H^2\dot\phi^2/M^2$ in Eqs. \eqref{eq2} and \eqref{eq3}
not only enhances the expansion rate, but also the friction of the motion of scalar field resulting slower roll of the inflaton. In the limit $M\rightarrow \infty$, the effect of the non-minimally derivative coupling is negligible,
Eqs. (\ref{eq2}) and (\ref{eq3}) reduce to the Friedmann equations in GR.
In this paper, we take the high friction limit $F=H^2/M^2\gg 1$ to take advantage of the gravitationally enhanced friction introduced by the non-minimally derivative coupling.

Under the slow-roll conditions,
\begin{equation}
\label{slrcond1}
\begin{split}
\frac{1}{2}\left(1+\frac{9H^2}{M^2}\right)\dot\phi^2\ll V(\phi),\\
|\ddot \phi|\ll |3H\dot\phi|,\\
\left|\frac{2\dot H}{M^2+3H^2}\right|\ll 1,
\end{split}
\end{equation}
Eqs. \eqref{eq2} and \eqref{eq3} are approximated as
\begin{gather}
\label{eq9}
H^2 \approx \frac{V(\phi)}{3},\\
\label{eq10}
3 H\dot\phi \left(1+\frac{3H^2}{M^2}\right)  \approx  - V_\phi,
\end{gather}
where $V_\phi=dV/d\phi$. To quantify those slow-roll conditions \eqref{slrcond1} in the high friction limit $H^2 \gg M^2$,
we introduce the following slow-roll parameters \cite{Germani:2010ux,Yang:2015pga}
\begin{gather}
\label{eqepsi}
\epsilon=\frac{1}{6}\left(\frac{V_\phi}{V} \right)^2\frac{1+9F}{(1+3F)^2}\approx\frac{M^2}{2V}\left(\frac{V_\phi}{V} \right)^2,\\
\label{eqeta}
\eta = \frac{1}{1+3F}\frac{V_{\phi\phi}}{V} \approx \frac{M^2}{V}\frac{V_{\phi\phi}}{V},
\end{gather}
where the approximation is valid in the high friction limit $F\gg 1$, and Eq. \eqref{eq9} was used to derive the approximation.
The scalar and tensor power spectra are \cite{Yang:2015pga}
\begin{equation}
\label{hfpseq1}
P_\zeta \approx \frac{1}{2\, \epsilon}\left(\frac{H}{2\pi}\right)^2
\left( \frac{c_sk}{aH} \right)^{3 - 2\nu },
\end{equation}
\begin{equation}
\label{hfpteq1}
P_T\approx 8\left(\frac{H}{2\pi}\right)^2\left( \frac{c_t k}{aH} \right)^{3 - 2\mu },
\end{equation}
where $c_s$ and $c_t$ are the effective sound speed, $\nu\approx 3/2+4\epsilon-\eta$ and $\mu=3/2+\epsilon$.
The amplitude of scalar power spectrum, the scalar spectral tilt and the tensor-to-scalar ratio are \cite{Germani:2010ux,Yang:2015pga}
\begin{equation}
\label{deras}
A_s\approx \frac{1}{8\pi^2}\frac{H^2}{\epsilon}\approx \frac{1}{24\pi^2}\frac{V(\phi)}{\epsilon},
\end{equation}
\begin{equation}
\label{eqns}
n_s =1+\left. \frac{d\ln P_\zeta }{d\ln k} \right|_{c_sk = aH}\approx 1 - 8\epsilon  + 2\eta,
\end{equation}
\begin{equation}
\label{eqr}
r =\frac{P_T}{P_\zeta}\approx 16\epsilon.
\end{equation}

In GR, the slow-roll parameters are defined as
\begin{gather}
\label{eqepsigr}
\epsilon_v=\frac{1}{2}\left(\frac{V_\phi}{V} \right)^2,\\
\label{eqetagr}
\eta_v = \frac{V_{\phi\phi}}{V}.
\end{gather}
The scalar spectral tilt and the tensor-to-scalar ratio are
\begin{equation}
\label{eqnsgr}
n_s \approx 1 - 6\epsilon_v  + 2\eta_v,
\end{equation}
\begin{equation}
\label{eqrgr}
r \approx 16\epsilon_v.
\end{equation}

The number of e-folds before the end of inflation is
\begin{equation}
\label{lythbld}
N(\phi_*)=\int_{\phi_*}^{\phi_e}\frac{H}{\dot\phi}d\phi=\int_{\phi_*}^{\phi_e}\frac{1}{\sqrt{2\epsilon}}\sqrt{\frac{3H^2}{M^2}}d\phi,
\end{equation}
where the subscript ``$e$" means the value of inflaton field at the end of inflation and the subscript ``$*$" means the value at the horizon crossing.
The value of $N$ is around $\sim 50 - 60$.

\section{Observational constraints}
\label{sec3}
In this section, we show how the models with non-minimally derivative coupling match the latest data release P-ACT-LB-BK18.
We consider the power law potential, the hilltop model, the polynomial model and E/T-models in the high friction limit.

\subsection{Power law potential}
\begin{figure}[htbp]
\centering
\includegraphics[width=0.6\textwidth]{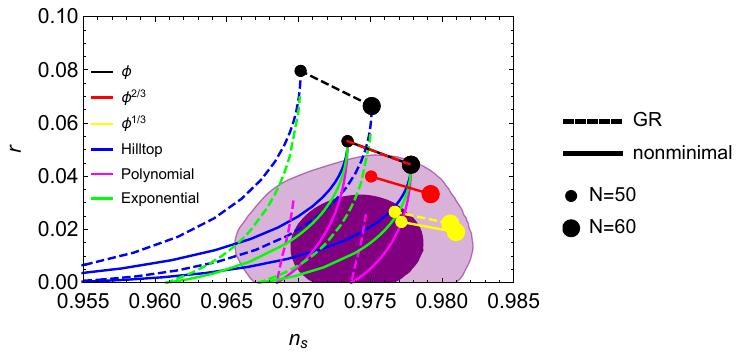}
\caption{Observational constraints on $n_s-r$ for different inflationary models with non-minimally derivative coupling.
The $1\sigma$ and $2\sigma$ contours are the observational results from the P-ACT-LB-BK18 data.
Solid and dashed lines represent predictions for the non-minimally derivative coupling case and GR, respectively,
with different colors distinguishing individual models.}
\label{nsr}
\end{figure}

For the chaotic inflation with the power law potential $V(\phi)=V_0\phi^n$ \cite{Linde:1983gd}, the slow-roll parameters are \cite{Yang:2015pga}
\begin{equation}
\label{pwrslreq1}
\begin{split}
\epsilon(\phi)&=\frac{n^2}{2V_0}\frac{M^2}{\phi^{n+2}},\\
\eta(\phi)&=\frac{n(n-1)}{V_0}\frac{M^2}{\phi^{n+2}}=\frac{2(n-1)}{n}\epsilon.
\end{split}
\end{equation}
Note that $M^2 \ll V_0$ in the high friction limit, without loss of generality, we take $M^2/V_0=10^{-4}$ in this paper.
The end of inflation is determined by ${\rm max}\{\epsilon(\phi_e),|\eta(\phi_e)|\}=1$.
For $0<n<2$, $|\eta|<\epsilon$, inflation ends when $\epsilon(\phi_e)=1$, so we have
\begin{equation}
\label{pwrslreq2}
\phi_e=\left(\frac{n^2M^2}{2V_0}\right)^{1/(n+2)}.
\end{equation}
The number of e-folds before the end of inflation is
\begin{equation}
\label{pwrslreq3a}
N=\frac{V_0\phi_*^{n+2}}{n(n+2)M^2 }-\frac{n}{2(n+2)}.
\end{equation}
For $n\ge 2$, $|\eta|>\epsilon$, inflation ends when $\eta(\phi_e)=1$, so
\begin{equation}
\label{pwrslreq2a}
\phi_e=\left(\frac{n(n-1)M^2}{V_0}\right)^{1/(n+2)},
\end{equation}
and the number of e-folds before the end of inflation is
\begin{equation}
\label{pwrslreq3b}
N=\frac{V_0\phi_*^{n+2}}{n(n+2)M^2 }-\frac{n-1}{n+2}.
\end{equation}
The value of the scalar field at the horizon exit is
\begin{equation}
\label{pwrslreq6}
\phi_*=\left(\frac{n(n+2)M^2\tilde N}{V_0}\right)^{1/(n+2)},
\end{equation}
where $\tilde N=N+n/2(n+2)$ for $0<n<2$ and $\tilde N=N+(n-1)/(n+2)$ for $n\ge 2$.
Substituting Eqs. \eqref{pwrslreq3a} and \eqref{pwrslreq3b} into Eqs. \eqref{eqns} and \eqref{eqr},
we obtain the scalar spectral tilt and the tensor-to-scalar ratio,
\begin{equation}
\label{plns}
n_s  \approx 1 - \frac{2(n+1)}{(n+2)\tilde N},
\end{equation}
\begin{equation}
\label{plr}
r \approx \frac{8n}{(n+2)\tilde N}.
\end{equation}
For $n=1$ and $N=50$, we get $n_s=0.973$ and $r=0.053$;
for $n=2/3$ and $N=50$, we get $n_s=0.975$ and $r=0.04$;
for $n=1/3$ and $N=50$, we get $n_s=0.977$ and $r=0.023$.
For $n=1$ and $N=60$, we get $n_s=0.978$ and $r=0.044$;
for $n=2/3$ and $N=60$, we get $n_s=0.979$ and $r=0.033$; for $n=1/3$ and $N=60$, we get $n_s=0.981$ and $r=0.019$.
The results are shown in Fig. \ref{nsr}. We also show the GR results \cite{Gao:2014ryw} in Fig. \ref{nsr}.
With the help of non-minimally derivative coupling, $n_s$ becomes larger and $r$ becomes smaller.
In particular, the linear potential with $n=1$ for $N=60$ is now marginally consistent with the observation at the $2\sigma$ confidence level.
Only the case with $n=1/3$ and $N=50$ is consistent with the $1\sigma$ contour for both non-minimally derivative coupling scenario and GR.

\subsection{Hilltop model}
For the hilltop model with the potential \cite{Boubekeur:2005zm}
\begin{equation}
\label{hilltopeq1}
V(\phi)=V_0\left(1-\frac{\phi^p}{\mu^p}\right),
\end{equation}
the slow-roll parameters in the high friction limit are \cite{Yang:2015pga}
\begin{equation}
\label{slrhilltopeq1}
\begin{split}
\epsilon(\phi)&=\frac{ p^2 M^2 (\phi/\mu)^{2 p-2}}{2 V_0 \mu^2 \left[1-(\phi/\mu)^p\right]^3},\\
\eta(\phi)&=-\frac{ p (p-1) M^2 (\phi/\mu)^{p-2}}{V_0 \mu^2 \left[1-(\phi/\mu)^p\right]^2}
=-\frac{2(p-1)}{p}\frac{1-(\phi/\mu)^p}{(\phi/\mu)^p}\epsilon(\phi).
\end{split}
\end{equation}

Since $\phi<\mu$, so $\epsilon(\phi)$ is smaller than $\eta(\phi)$ by a factor $(\phi/\mu)^p$ when $p\ge 3$.
In this model, we take $p=4$ as a representative case.
By using Eqs. \eqref{eqns}, \eqref{eqr} and \eqref{lythbld},
we numerically calculate the observables $n_s$ and $r$ with $\mu$ varying
and the results are shown in Fig. \ref{nsr}.
The solid blue lines represent the results for hilltop model with $p=4$ in the high friction limit,
where the left and the right lines correspond to $N=50$ and $N=60$, respectively.
In Fig. \ref{nsr}, we also show the results in GR with dashed blue lines.
As seen from Fig. \ref{nsr}, the non-minimally derivative coupling increases $n_s$ and suppresses $r$, aligning the model with current observational constraints.
For $N=50$ the model can be consistent with the P-ACT-LB-BK18 data only at the $2\sigma$ confidence level,
while satisfying current observations at the $1\sigma$ confidence level for $N=60$.
For $N=50$, we find that $0.18<\mu<2.4$ at the $2\sigma$ confidence level.
For $N=60$, we find that $0.165<\mu<0.69$ at the $1\sigma$ confidence level and $\mu>0.123$ at the $2\sigma$ confidence level.

From Fig. \ref{nsr}, we see that in the large $\mu$ limit the model reaches an attractor corresponding to the results for the linear potential.
The attractor behavior can be understood through the following analysis.
In the limit of $\mu\gg1$, we get $\phi_e\sim \mu$ and $\phi_*\sim \mu$.
Taylor expanding the potential \eqref{hilltopeq1} to the first order, we get the linear potential $V(\phi)\sim \phi$.
Therefore, the attractors are given by Eqs. \eqref{plns} and  \eqref{plr} with $n=1$ which is independent of the parameter $p$.
Specifically, $n_s\approx0.973$ and $r\approx0.053$ for $N=50$, and $n_s\approx0.978$ and $r\approx0.044$ for $N=60$.

As discussed above, $\epsilon(\phi)$ is smaller than $\eta(\phi)$ by a factor $(\phi/\mu)^p$ when $p\ge 3$, so $\epsilon(\phi)<|\eta(\phi)|$, inflation ends when $|\eta(\phi_e)|=1$.
In the limit of $\mu\ll1$, we get small field inflation, $\phi\ll \mu$,
\begin{gather}
\label{phiehilltopeq2}
\left(\frac{\phi_e}{\mu}\right)^{p-2}\approx \frac{V_0\mu^2}{p(p-1)M^2}.
\end{gather}
The number of e-folds before the end of inflation when the pivotal scale $k_*$ crosses the horizon is $N=f(\phi_e/\mu)-f(\phi_*/\mu)$, where
\begin{equation}
\label{nhilltopeq1}
f(x)=\frac{V_0 \mu ^2}{p M^2 }\left(\frac{x^{-p}}{2-p}+\frac{x^p}{p+2}-1\right)x^2,
\end{equation}
for $p\neq 2$. In the limit of $\phi\ll \mu$, we get
\begin{equation}
\label{nhilltopeq2}
N\approx \frac{V_0 \mu ^2}{p(p-2) M^2 }\left(\frac{\mu}{\phi_*}\right)^{p-2}-\frac{p-1}{p-2}.
\end{equation}
Therefore, the scalar spectral index is
\begin{equation}
\label{nshilltopeq3}
n_s\approx 1-\frac{2(p-1)}{(p-2)[N+(p-1)/(p-2)]},
\end{equation}
and the tensor-to-scalar ratio is
\begin{equation}
\label{rhilltopeq3}
r\approx \frac{8px_*^p}{(p-2)[N+(p-1)/(p-2)]}\ll 1,
\end{equation}
where $x_*=\phi_*/\mu$.
For $N=50$ and $p=4$, we get $n_s\approx0.942$ and $r\approx1.83\times 10^{-10}$.
For $N=60$ and $p=4$, we get $n_s\approx0.951$ and $r\approx1.07\times 10^{-10}$.
These analytical estimations agree with the numerical results.

In GR, the slow-roll parameters are
\begin{equation}
\label{grht1}
\begin{split}
{\color{blue}\epsilon_v}(\phi)&=\frac{ p^2 (\phi/\mu)^{2 p-2}}{2 \mu^2 \left[1-(\phi/\mu)^p\right]^2},\\
{\color{blue}\eta_v}(\phi)&=-\frac{ p (p-1) (\phi/\mu)^{p-2}}{\mu^2 \left[1-(\phi/\mu)^p\right]}
=-\frac{2(p-1)}{p}\frac{1-(\phi/\mu)^p}{(\phi/\mu)^p}{\color{blue}\epsilon_v}(\phi).
\end{split}
\end{equation}
When $p\ge 3$, $\epsilon_v(\phi)<|\eta_v(\phi)|$, the end of inflation is determined by $|{\color{blue}\eta_v}(\phi_e)|=1$ .
The number of e-folds before the end of inflation is $N=g(\phi_e/\mu)-g(\phi_*/\mu)$, where
\begin{equation}
\label{grht2}
g(x)=\frac{\mu ^2}{p}\left(\frac{x^{2-p}}{2-p}-\frac{1}{2}x^2\right),
\end{equation}
for $p\neq 2$. Again in the limit of $\mu\gg1$, $\phi\sim \mu$ and the potential behaves as the linear potential $V\sim \phi$,
i.e., $n_s \approx 0.97$ and $r\approx0.078$ for $N=50$; $n_s\approx0.975$ and $r\approx0.065$ for $N=60$.
When $\mu\ll 1$, the attractors are the same as those in non-minimally derivative coupling case given by  Eqs. \eqref{nshilltopeq3} and \eqref{rhilltopeq3}.

\subsection{Polynomial $\alpha$-attractors}
For the polynomial $\alpha$-attractors \cite{Kallosh:2022feu}, the potential is
\begin{equation}
\label{polyv}
V(\phi)=V_0\left(1-\frac{\mu^p}{\phi^p}\right).
\end{equation}
The model is also called D-brane inflation \cite{Kallosh:2019jnl}.
The slow-roll parameters in the high friction limit are
\begin{equation}
\label{poly1}
\begin{split}
\epsilon(\phi)&=\frac{ p^2 M^2 (\mu/\phi)^{2 p+2}}{2 V_0 \mu^2 \left[1-(\mu/\phi)^p\right]^3},\\
\eta(\phi)&=-\frac{ p (p+1) M^2 (\mu/\phi)^{p+2}}{V_0 \mu^2 \left[1-(\mu/\phi)^p\right]^2}
=-\frac{2(p+1)}{p}\frac{1-(\mu/\phi)^p}{(\mu/\phi)^p}\epsilon(\phi).
\end{split}
\end{equation}
Without loss of generality, we take $p=3$.
By using Eqs. \eqref{eqns}, \eqref{eqr} and \eqref{lythbld},
we numerically calculate the observables $n_s$ and $r$  with $\mu$ varying, and the results are shown in Fig. \ref{nsr} with solid magenta lines.
With the help of non-minimally derivative coupling, the polynomial $\alpha$-attractors are consistent with the P-ACT-LB-BK18 data for
both $N=50$ and $N=60$.
For $N=50$, the $1\sigma$ and $2\sigma$ constraints on $\mu$ by the  P-ACT-LB-BK18 data are $0.032<\mu<0.51$ and $0<\mu<3.56$, respectively;
while for $N=60$, the $1\sigma$ constraint is $0<\mu<0.57$ and at the $2\sigma$ confidence level, $\mu$ can be arbitrarily large.

In the limit of  $\mu\gg1$, the same attractors \eqref{plns} and  \eqref{plr} with $n=1$ are reached because the potential behaves as the linear potential when $\phi\sim \mu$.
On the other hand, when $\mu\ll 1$, we have small field inflation.
From Eq. \eqref{poly1}, we see that $\epsilon(\phi)<|\eta(\phi)|$, so $|\eta(\phi_e)|=1$ and we get
\begin{equation}
\label{phiepoly2}
\left(\frac{\mu}{\phi_e}\right)^{p+2}\approx \frac{V_0\mu^2}{p(p+1)M^2},
\end{equation}
and $N=h(\mu/\phi_e)-h(\mu/\phi_*)$, where
\begin{equation}
\label{npoly1}
h(x)=\frac{V_0 \mu ^2}{p M^2 }\left(\frac{x^{-p}}{2+p}+\frac{x^p}{2-p}-1\right)x^{-2},
\end{equation}
for $p\neq 2$. Combining Eqs. \eqref{phiepoly2} and \eqref{npoly1}, we get
\begin{equation}
\label{npoly2}
N\approx \frac{V_0 \mu ^2}{p(p+2) M^2 }\left(\frac{\mu}{\phi_*}\right)^{-p-2}-\frac{p+1}{p+2},
\end{equation}
so
\begin{equation}
\label{nspoly}
n_s\approx 1-\frac{2(p+1)}{(p+2)[N+(p+1)/(p+2)]},
\end{equation}
and
\begin{equation}
\label{rpoly}
r\approx \frac{8px_*^p}{(p+2)[N+(p+1)/(p+2)]}\ll 1.
\end{equation}
For $p=3$, the attractors are $n_s\approx0.969$ and $r\approx4.43\times 10^{-7}$ for $N=50$,
and $n_s\approx0.974$ and $r\approx3.32\times 10^{-7}$ for $N=60$, these analytical results \eqref{nspoly} and \eqref{rpoly} are consistent with the numerical results.

In GR, the slow-roll parameters are
\begin{equation}
\label{grht1}
\begin{split}
{\color{blue}\epsilon_v}(\phi)&=\frac{ p^2 (\mu/\phi)^{2 p+2}}{2 \mu^2 \left[1-(\mu/\phi)^p\right]^2},\\
{\color{blue}\eta_v}(\phi)&=-\frac{ p (p+1) (\mu/\phi)^{p+2}}{\mu^2 \left[1-(\mu/\phi)^p\right]}
=-\frac{2(p+1)}{p}\frac{1-(\mu/\phi)^p}{(\mu/\phi)^p}{\color{blue}\epsilon_v}(\phi).
\end{split}
\end{equation}
The number of e-folds before the end of inflation is $N=k(\mu/\phi_e)-k(\mu/\phi_*)$, where
\begin{equation}
\label{grht2}
k(x)=\frac{\mu ^2}{p}\left(\frac{x^{-2-p}}{2+p}-\frac{1}{2}x^{-2}\right).
\end{equation}
The results for $n_s$ and $r$ with $p=3$ in GR are also shown in Fig. \ref{nsr} as dashed magenta lines.
It is evident that the results for $N=50$ are excluded by observational data at the $1\sigma$ confidence level,
while those for $N=60$ are consistent with the data at the $1\sigma$ level.
The same attractors \eqref{nspoly} and \eqref{rpoly} as in the non-minimally derivative coupling case are obtained in the limit $\mu\ll 1$.

\subsection{Exponential $\alpha$-attractors}
The E model and T model share the same asymptotic form of the potential in the large field limit \cite{Kallosh:2013hoa,Yi:2016jqr},
\begin{equation}
\label{nonemv}
V(\phi)=V_0\left[1-a\exp\left(-\sqrt{\frac{2}{3\alpha}}\phi\right)\right].
\end{equation}
These models were also called exponential $\alpha$-attractors in \cite{Kallosh:2022feu}.
In this section, without loss of generality we consider $a=1$ for simplicity. The slow-roll parameters are
\begin{equation}
\label{nonem1}
\begin{split}
\epsilon(\phi)&=\frac{M^2 \exp\left(\sqrt{\frac{2}{3\alpha}}\phi\right)}{3\alpha V_0\left[\exp\left(\sqrt{\frac{2}{3\alpha}}\phi\right)-1\right]^3},\\
\eta(\phi)&=-\frac{2M^2 \exp\left(\sqrt{\frac{2}{3\alpha}}\phi\right)}{3\alpha V_0\left[\exp\left(\sqrt{\frac{2}{3\alpha}}\phi\right)-1\right]^2}.
\end{split}
\end{equation}
The number of e-folds before the end of inflation is
\begin{equation}
N=m\left(\sqrt{\frac{2}{3\alpha}}\phi_e\right)-m\left(\sqrt{\frac{2}{3\alpha}}\phi_*\right),
\end{equation}
where
\begin{equation}
\label{nonemn}
m(x)=\frac{V_0}{2 M^2 }\left[3\alpha\exp(x)-3\alpha\exp(-x)-6\alpha x\right].
\end{equation}
By using Eqs. \eqref{eqns}, \eqref{eqr} and \eqref{lythbld},
we numerically calculate $n_s$ and $r$ with varying $\alpha$ and the results are shown in Fig. \ref{nsr}.
It is evident that the model is consistent with the P-ACT-LB-BK18 data at the $2\sigma$ confidence level with the help of non-minimally derivative coupling.
For $N=50$, the $1\sigma$ and $2\sigma$ constraints on $\alpha$ are $1.2\times10^{-3}<\alpha<0.011$ and $2.5\times10^{-4}<\alpha<0.468$, respectively.
For $N=60$, the $1\sigma$ and $2\sigma$ constraints are $1.38\times10^{-4}<\alpha<0.022$ and $\alpha>0$, respectively.

To understand the attractor behavior, we consider the limits of large and small $\alpha$.
In the large $\alpha$ limit, $\sqrt{2/(3\alpha)}\,\phi\ll 1$. To first order of $\sqrt{2/(3\alpha)}\,\phi$, the potential can be approximated as linear.
As shown in Fig. \ref{nsr}, the results \eqref{plns} and \eqref{plr} with $n=1$ give the upper limits for the exponential $\alpha$-attractors.

In the small $\alpha$ limit,
$\alpha\ll1$, the end of inflation is determined by $|\eta(\phi_e)|=1$ because $\epsilon(\phi)<|\eta(\phi)|$, so we have
\begin{gather}
\label{phienonem}
\exp\left(\sqrt{\frac{2}{3\alpha}}\phi_e\right)\approx \frac{2M^2}{3\alpha V_0},
\end{gather}
and
\begin{equation}
\label{nonemn1}
N\approx \frac{3\alpha V_0}{2 M^2 }\exp\left(\sqrt{\frac{2}{3\alpha}}\phi_*\right)-1.
\end{equation}
So
\begin{equation}
\label{nsem}
n_s\approx 1-\frac{2}{N+1},
\end{equation}
and
\begin{equation}
\label{rem}
r\approx \frac{12\alpha V_0}{M^2 N^2}\ll 1.
\end{equation}
These analytical attractors in Eqs. \eqref{nsem} and \eqref{rem} yield $n_s\approx0.961$ and $r\approx4.61\times 10^{-7}$ for $N=50$;
$n_s\approx0.967$ and $r\approx3.22\times 10^{-7}$ for $N=60$,
which are consistent with the numerical results shown in Fig. \ref{nsr}.

In GR, the slow-roll parameters are
\begin{equation}
\label{grem1}
\begin{split}
{\color{blue}\epsilon_v}(\phi)&=\frac{ 1}{3\alpha \left[\exp\left(\sqrt{\frac{2}{3\alpha}}\phi\right)-1\right]^2},\\
{\color{blue}\eta_v}(\phi)&=\frac{2}{3\alpha\left[1-\exp\left(\sqrt{\frac{2}{3\alpha}}\phi\right)\right]}.
\end{split}
\end{equation}
The number of e-folds before the end of inflation is
\begin{equation}
N=l\left(\sqrt{\frac{2}{3\alpha}}\phi_e\right)-l\left(\sqrt{\frac{2}{3\alpha}}\phi_*\right),
\end{equation}
where
\begin{equation}
\label{gremn}
l(x)=\frac{3\alpha \exp(x)}{2}-\frac{3\alpha x}{2}.
\end{equation}
The results for $n_s$ and $r$ of exponential $\alpha$-attractors in GR are shown in Fig. \ref{nsr} as dashed green lines.
We see that the case with $N=50$ is ruled out by P-ACT-LB-BK18 data at more than the $2\sigma$ level, while the case with $N=60$ is consistent with the data at the $1\sigma$ level.
In the small $\alpha$ limit, the attractor \eqref{nsem}  for $n_s$ is obtained, but the attractor for the tensor-to-scalar ratio is $r=12\alpha/N^2$ which is smaller than that with non-minimally derivative coupling by a factor of $V_0/M^2$.

\section{Conclusions}
\label{sec4}
For models with non-minimal derivative coupling in the high friction limit,
we calculate the scalar spectral index $n_s$ and the tensor-to-scalar ratio $r$.
We then analyze the numerical results for both the non-minimally derivative coupling and GR cases, deriving parameter constraints based on the P-ACT-LB-BK18 data.

For the power law potential, the presence of non-minimally derivative coupling enables the inflaton field excursion to be sub-Planckian.
Specifically, the linear potential with $n=1$ and $N=60$ is now marginally consistent with the observations at the $2\sigma$ confidence level.
Meanwhile, the case with $n=1/3$ and $N=50$ lies well within the $1\sigma$ contour for both non-minimally derivative coupling scenario and GR.

For the hilltop model, the scale $\mu$ can be smaller than the Planck energy, allowing for small-field inflation.
In the case $p=4$, the scenario with $N=50$ lies within the $2\sigma$ $n_s-r$ contour from the P-ACT-LB-BK18 data,
while the case with $N=60$ falls within both the $1\sigma$ and $2\sigma$ $n_s-r$ contours.
For $N=50$, we constrain the scale as $0.18<\mu<2.4$ at the $2\sigma$ confidence level.
For $N=60$, the bounds are $0.165<\mu<0.69$ at $1\sigma$, with a lower limit $\mu>0.123$ at $2\sigma$.
Additionally, we derive analytical expressions for attractor behaviors in the limits of large and small $\mu$,
for both the non-minimal derivative coupling scenario and GR.

For the polynomial $\alpha$-attractors with $p=3$, both the $N=50$ and $N=60$ cases are consistent with the P-ACT-LB-BK18 data.
For $N=50$, the $1\sigma$ and $2\sigma$ constraints on $\mu$ are $0.032<\mu<0.51$ and $0<\mu<3.56$, respectively.
For $N=60$, the $1\sigma$ constraint is $0<\mu<0.57$, while at the $2\sigma$ level, $\mu$ is unconstrained and can be arbitrarily large.
The numerical results confirm the analytical attractor expressions for both the non-minimally derivative coupling scenario and GR in the limits of large and small $\mu$.

For the exponential $\alpha$-attractors with $a=1$, both the $N=50$ and $N=60$ cases are consistent with the P-ACT-LB-BK18 data at the $1\sigma$ confidence level.
For $N=50$, the $1\sigma$ and $2\sigma$ constraints on $\alpha$ are $1.259\times10^{-9}<\alpha<1.072\times10^{-8}$ and $2.512\times10^{-10}<\alpha<3.981\times10^{-7}$, respectively.
For $N=60$, the $1\sigma$ constraint is $1.413\times10^{-10}<\alpha<2.239\times10^{-8}$,
while at the $2\sigma$ level,
only the condition
and only $\alpha>0$ is required at $2\sigma$.
The numerical results confirm the analytical attractor expressions for both the non-minimally derivative coupling scenario and GR in the limits of large and small $\alpha$.

In conclusion, the non-minimally kinetic coupling to the Einstein tensor effectively lowers the energy scale of the models to sub-Planckian levels, enabling small-field inflation within these frameworks.
This non-minimally derivative coupling allows a broader class of inflationary models to be consistent with the latest observational data P-ACT-LB-BK18.

\begin{acknowledgments}
This work is supported in part by the National Natural Science Foundation of China under Grant Nos. 12175184 and 12205015, the National Key Research and Development Program of China under Grant No. 2020YFC2201504, the supporting fund for Young Researcher of Beijing Normal University under Grant No. 28719/310432102, the Chongqing Natural Science Foundation under Grant No. CSTB2022NSCQ-MSX1324.

\end{acknowledgments}


\begin{thebibliography}{10}

\bibitem{Louis:2025tst}
{\scshape ACT} collaboration, T.~Louis et~al., \emph{{The Atacama Cosmology Telescope: DR6 Power Spectra, Likelihoods and $\Lambda$CDM Parameters}},  \href{https://arxiv.org/abs/2503.14452}{{\ttfamily 2503.14452}}.

\bibitem{ACT:2025tim}
{\scshape ACT} collaboration, E.~Calabrese et~al., \emph{{The Atacama Cosmology Telescope: DR6 Constraints on Extended Cosmological Models}},  \href{https://arxiv.org/abs/2503.14454}{{\ttfamily 2503.14454}}.

\bibitem{Starobinsky:1980te}
A.~A. Starobinsky, \emph{{A New Type of Isotropic Cosmological Models Without Singularity}}, \href{https://doi.org/10.1016/0370-2693(80)90670-X}{\emph{Phys. Lett. B} {\bfseries 91} (1980) 99--102}.

\bibitem{Kaiser:1994vs}
D.~I. Kaiser, \emph{{Primordial spectral indices from generalized Einstein theories}}, \href{https://doi.org/10.1103/PhysRevD.52.4295}{\emph{Phys. Rev. D} {\bfseries 52} (1995) 4295--4306}, [\href{https://arxiv.org/abs/astro-ph/9408044}{{\ttfamily astro-ph/9408044}}].

\bibitem{Bezrukov:2007ep}
F.~L. Bezrukov and M.~Shaposhnikov, \emph{{The Standard Model Higgs boson as the inflaton}}, \href{https://doi.org/10.1016/j.physletb.2007.11.072}{\emph{Phys. Lett. B} {\bfseries 659} (2008) 703--706}, [\href{https://arxiv.org/abs/0710.3755}{{\ttfamily 0710.3755}}].

\bibitem{Kallosh:2013hoa}
R.~Kallosh and A.~Linde, \emph{{Universality Class in Conformal Inflation}}, \href{https://doi.org/10.1088/1475-7516/2013/07/002}{\emph{JCAP} {\bfseries 07} (2013) 002}, [\href{https://arxiv.org/abs/1306.5220}{{\ttfamily 1306.5220}}].

\bibitem{Kallosh:2013maa}
R.~Kallosh and A.~Linde, \emph{{Non-minimal Inflationary Attractors}}, \href{https://doi.org/10.1088/1475-7516/2013/10/033}{\emph{JCAP} {\bfseries 10} (2013) 033}, [\href{https://arxiv.org/abs/1307.7938}{{\ttfamily 1307.7938}}].

\bibitem{Lin:2015fqa}
J.~Lin, Q.~Gao and Y.~Gong, \emph{{The reconstruction of inflationary potentials}}, \href{https://doi.org/10.1093/mnras/stw915}{\emph{Mon. Not. Roy. Astron. Soc.} {\bfseries 459} (2016) 4029--4037}, [\href{https://arxiv.org/abs/1508.07145}{{\ttfamily 1508.07145}}].

\bibitem{Yi:2016jqr}
Z.~Yi and Y.~Gong, \emph{{Nonminimal coupling and inflationary attractors}}, \href{https://doi.org/10.1103/PhysRevD.94.103527}{\emph{Phys. Rev. D} {\bfseries 94} (2016) 103527}, [\href{https://arxiv.org/abs/1608.05922}{{\ttfamily 1608.05922}}].

\bibitem{Fei:2017fub}
Q.~Fei, Y.~Gong, J.~Lin and Z.~Yi, \emph{{The reconstruction of tachyon inflationary potentials}}, \href{https://doi.org/10.1088/1475-7516/2017/08/018}{\emph{JCAP} {\bfseries 08} (2017) 018}, [\href{https://arxiv.org/abs/1705.02545}{{\ttfamily 1705.02545}}].

\bibitem{Gao:2017uja}
Q.~Gao and Y.~Gong, \emph{{Reconstruction of extended inflationary potentials for attractors}}, \href{https://doi.org/10.1140/epjp/i2018-12324-3}{\emph{Eur. Phys. J. Plus} {\bfseries 133} (2018) 491}, [\href{https://arxiv.org/abs/1703.02220}{{\ttfamily 1703.02220}}].

\bibitem{Planck:2018jri}
{\scshape Planck} collaboration, Y.~Akrami et~al., \emph{{Planck 2018 results. X. Constraints on inflation}}, \href{https://doi.org/10.1051/0004-6361/201833887}{\emph{Astron. Astrophys.} {\bfseries 641} (2020) A10}, [\href{https://arxiv.org/abs/1807.06211}{{\ttfamily 1807.06211}}].

\bibitem{Planck:2018vyg}
{\scshape Planck} collaboration, N.~Aghanim et~al., \emph{{Planck 2018 results. VI. Cosmological parameters}}, \href{https://doi.org/10.1051/0004-6361/201833910}{\emph{Astron. Astrophys.} {\bfseries 641} (2020) A6}, [\href{https://arxiv.org/abs/1807.06209}{{\ttfamily 1807.06209}}].

\bibitem{DESI:2024uvr}
{\scshape DESI} collaboration, A.~G. Adame et~al., \emph{{DESI 2024 III: baryon acoustic oscillations from galaxies and quasars}}, \href{https://doi.org/10.1088/1475-7516/2025/04/012}{\emph{JCAP} {\bfseries 04} (2025) 012}, [\href{https://arxiv.org/abs/2404.03000}{{\ttfamily 2404.03000}}].

\bibitem{DESI:2024mwx}
{\scshape DESI} collaboration, A.~G. Adame et~al., \emph{{DESI 2024 VI: cosmological constraints from the measurements of baryon acoustic oscillations}}, \href{https://doi.org/10.1088/1475-7516/2025/02/021}{\emph{JCAP} {\bfseries 02} (2025) 021}, [\href{https://arxiv.org/abs/2404.03002}{{\ttfamily 2404.03002}}].

\bibitem{BICEP:2021xfz}
{\scshape BICEP, Keck} collaboration, P.~A.~R. Ade et~al., \emph{{Improved Constraints on Primordial Gravitational Waves using Planck, WMAP, and BICEP/Keck Observations through the 2018 Observing Season}}, \href{https://doi.org/10.1103/PhysRevLett.127.151301}{\emph{Phys. Rev. Lett.} {\bfseries 127} (2021) 151301}, [\href{https://arxiv.org/abs/2110.00483}{{\ttfamily 2110.00483}}].

\bibitem{Gialamas:2025kef}
I.~D. Gialamas, A.~Karam, A.~Racioppi and M.~Raidal, \emph{{Has ACT measured radiative corrections to the tree-level Higgs-like inflation?}},  \href{https://arxiv.org/abs/2504.06002}{{\ttfamily 2504.06002}}.

\bibitem{Frob:2025sfq}
M.~B. Fr{\"o}b, D.~Glavan, P.~Meda and I.~Sawicki, \emph{{One-loop correction to primordial tensor modes during radiation era}},  \href{https://arxiv.org/abs/2504.02609}{{\ttfamily 2504.02609}}.

\bibitem{Dioguardi:2025vci}
C.~Dioguardi, A.~J. Iovino and A.~Racioppi, \emph{{Fractional attractors in light of the latest ACT observations}}, \href{https://doi.org/10.1016/j.physletb.2025.139664}{\emph{Phys. Lett. B} {\bfseries 868} (2025) 139664}, [\href{https://arxiv.org/abs/2504.02809}{{\ttfamily 2504.02809}}].

\bibitem{Brahma:2025dio}
S.~Brahma and J.~Calder{\'o}n-Figueroa, \emph{{Is the CMB revealing signs of pre-inflationary physics?}},  \href{https://arxiv.org/abs/2504.02746}{{\ttfamily 2504.02746}}.

\bibitem{Berera:2025vsu}
A.~Berera, S.~Brahma, Z.~Qiu, R.~O.~Ramos and G.~S. Rodrigues, \emph{{The early universe is $\textit{ACT}$-ing $\textit{warm}$}},  \href{https://arxiv.org/abs/2504.02655}{{\ttfamily 2504.02655}}.

\bibitem{Aoki:2025wld}
S.~Aoki, H.~Otsuka and R.~Yanagita, \emph{{Higgs-Modular Inflation}},  \href{https://arxiv.org/abs/2504.01622}{{\ttfamily 2504.01622}}.

\bibitem{Dioguardi:2025mpp}
C.~Dioguardi and A.~Karam, \emph{{Palatini linear attractors are back in action}}, \href{https://doi.org/10.1103/23b3-9d7q}{\emph{Phys. Rev. D} {\bfseries 111} (2025) 123521}, [\href{https://arxiv.org/abs/2504.12937}{{\ttfamily 2504.12937}}].

\bibitem{Salvio:2025izr}
A.~Salvio, \emph{{Independent connection in ACTion during inflation}},  \href{https://arxiv.org/abs/2504.10488}{{\ttfamily 2504.10488}}.

\bibitem{Liu:2025qca}
L.~Liu, Z.~Yi and Y.~Gong, \emph{{Reconciling Higgs Inflation with ACT Observations through Reheating}},  \href{https://arxiv.org/abs/2505.02407}{{\ttfamily 2505.02407}}.

\bibitem{Haque:2025uis}
M.~R. Haque, S.~Pal and D.~Paul, \emph{{Improved Predictions on Higgs-Starobinsky Inflation and Reheating with ACT DR6 and Primordial Gravitational Waves}},  \href{https://arxiv.org/abs/2505.04615}{{\ttfamily 2505.04615}}.

\bibitem{Zharov:2025evb}
D.~S. Zharov, O.~O. Sobol and S.~I. Vilchinskii, \emph{{Reheating ACTs on Starobinsky and Higgs inflation}},  \href{https://arxiv.org/abs/2505.01129}{{\ttfamily 2505.01129}}.

\bibitem{Haque:2025uri}
M.~R. Haque, S.~Pal and D.~Paul, \emph{{ACT DR6 Insights on the Inflationary Attractor models and Reheating}},  \href{https://arxiv.org/abs/2505.01517}{{\ttfamily 2505.01517}}.

\bibitem{Drees:2025ngb}
M.~Drees and Y.~Xu, \emph{{Refined predictions for Starobinsky inflation and post-inflationary constraints in light of ACT}}, \href{https://doi.org/10.1016/j.physletb.2025.139612}{\emph{Phys. Lett. B} {\bfseries 867} (2025) 139612}, [\href{https://arxiv.org/abs/2504.20757}{{\ttfamily 2504.20757}}].

\bibitem{He:2025bli}
M.~He, M.~Hong and K.~Mukaida, \emph{{Increase of $n_s$ in regularized pole inflation {\&} Einstein-Cartan gravity}},  \href{https://arxiv.org/abs/2504.16069}{{\ttfamily 2504.16069}}.

\bibitem{Yogesh:2025wak}
Yogesh, A.~Mohammadi, Q.~Wu and T.~Zhu, \emph{{Starobinsky like inflation and EGB Gravity in the light of ACT}},  \href{https://arxiv.org/abs/2505.05363}{{\ttfamily 2505.05363}}.

\bibitem{Gialamas:2025ofz}
I.~D. Gialamas, T.~Katsoulas and K.~Tamvakis, \emph{{Keeping the relation between the Starobinsky model and no-scale supergravity ACTive}},  \href{https://arxiv.org/abs/2505.03608}{{\ttfamily 2505.03608}}.

\bibitem{Peng:2025bws}
Z.-Z. Peng, Z.-C. Chen and L.~Liu, \emph{{The polynomial potential inflation in light of ACT observations}},  \href{https://arxiv.org/abs/2505.12816}{{\ttfamily 2505.12816}}.

\bibitem{Mondal:2025kur}
R.~Mondal, S.~Mondal and A.~Chakraborty, \emph{{Constraining Reheating Temperature, Inflaton-SM Coupling and Dark Matter Mass in Light of ACT DR6 Observations}},  \href{https://arxiv.org/abs/2505.13387}{{\ttfamily 2505.13387}}.

\bibitem{Yi:2025dms}
Z.~Yi, X.~Wang, Q.~Gao and Y.~Gong, \emph{{Potential Reconstruction from ACT Observations Leading to Polynomial $\alpha$-Attractor}},  \href{https://arxiv.org/abs/2505.10268}{{\ttfamily 2505.10268}}.

\bibitem{Addazi:2025qra}
A.~Addazi, Y.~Aldabergenov and S.~V. Ketov, \emph{{Curvature corrections to Starobinsky inflation can explain the ACT results}},  \href{https://arxiv.org/abs/2505.10305}{{\ttfamily 2505.10305}}.

\bibitem{Maity:2025czp}
S.~Maity, \emph{{ACT-ing on inflation: Implications of non Bunch-Davies initial condition and reheating on single-field slow roll models}},  \href{https://arxiv.org/abs/2505.10534}{{\ttfamily 2505.10534}}.

\bibitem{Byrnes:2025kit}
C.~T. Byrnes, M.~Cort{\^e}s and A.~R. Liddle, \emph{{The curvaton ACTs again}},  \href{https://arxiv.org/abs/2505.09682}{{\ttfamily 2505.09682}}.

\bibitem{Kallosh:2025rni}
R.~Kallosh, A.~Linde and D.~Roest, \emph{{A simple scenario for the last ACT}},  \href{https://arxiv.org/abs/2503.21030}{{\ttfamily 2503.21030}}.

\bibitem{Gao:2025onc}
Q.~Gao, Y.~Gong, Z.~Yi and F.~Zhang, \emph{{Non-minimal coupling in light of ACT}},  \href{https://arxiv.org/abs/2504.15218}{{\ttfamily 2504.15218}}.

\bibitem{Katsoulas:2025mcu}
T.~Katsoulas and K.~Tamvakis, \emph{{General Einstein-Cartan quadratic gravity with derivative couplings}}, \href{https://doi.org/10.1088/1475-7516/2025/06/022}{\emph{JCAP} {\bfseries 06} (2025) 022}, [\href{https://arxiv.org/abs/2502.16980}{{\ttfamily 2502.16980}}].

\bibitem{Wolf:2025ecy}
W.~J. Wolf, \emph{{Inflationary attractors and radiative corrections in light of ACT}},  \href{https://arxiv.org/abs/2506.12436}{{\ttfamily 2506.12436}}.

\bibitem{Odintsov:2025wai}
S.~D. Odintsov and V.~K. Oikonomou, \emph{{GW170817 Viable Einstein-Gauss-Bonnet Inflation Compatible with the Atacama Cosmology Telescope Data}},  \href{https://arxiv.org/abs/2506.08193}{{\ttfamily 2506.08193}}.

\bibitem{Pallis:2025nrv}
C.~Pallis, \emph{{Kinetically modified Palatini inflation meets ACT data}}, \href{https://doi.org/10.1016/j.physletb.2025.139739}{\emph{Phys. Lett. B} {\bfseries 868} (2025) 139739}, [\href{https://arxiv.org/abs/2505.23243}{{\ttfamily 2505.23243}}].

\bibitem{Guo:2010jr}
Z.-K. Guo and D.~J. Schwarz, \emph{{Slow-roll inflation with a Gauss-Bonnet correction}}, \href{https://doi.org/10.1103/PhysRevD.81.123520}{\emph{Phys. Rev. D} {\bfseries 81} (2010) 123520}, [\href{https://arxiv.org/abs/1001.1897}{{\ttfamily 1001.1897}}].

\bibitem{Jiang:2013gza}
P.-X. Jiang, J.-W. Hu and Z.-K. Guo, \emph{{Inflation coupled to a Gauss-Bonnet term}}, \href{https://doi.org/10.1103/PhysRevD.88.123508}{\emph{Phys. Rev. D} {\bfseries 88} (2013) 123508}, [\href{https://arxiv.org/abs/1310.5579}{{\ttfamily 1310.5579}}].

\bibitem{Boubekeur:2015xza}
L.~Boubekeur, E.~Giusarma, O.~Mena and H.~Ram{\'\i}rez, \emph{{Does Current Data Prefer a Non-minimally Coupled Inflaton?}}, \href{https://doi.org/10.1103/PhysRevD.91.103004}{\emph{Phys. Rev. D} {\bfseries 91} (2015) 103004}, [\href{https://arxiv.org/abs/1502.05193}{{\ttfamily 1502.05193}}].

\bibitem{Amendola:1993uh}
L.~Amendola, \emph{{Cosmology with nonminimal derivative couplings}}, \href{https://doi.org/10.1016/0370-2693(93)90685-B}{\emph{Phys. Lett. B} {\bfseries 301} (1993) 175--182}, [\href{https://arxiv.org/abs/gr-qc/9302010}{{\ttfamily gr-qc/9302010}}].

\bibitem{Boulware:1972zf}
D.~G. Boulware and S.~Deser, \emph{{Inconsistency of finite range gravitation}}, \href{https://doi.org/10.1016/0370-2693(72)90418-2}{\emph{Phys. Lett. B} {\bfseries 40} (1972) 227--229}.

\bibitem{Horndeski:1974wa}
G.~W. Horndeski, \emph{{Second-order scalar-tensor field equations in a four-dimensional space}}, \href{https://doi.org/10.1007/BF01807638}{\emph{Int. J. Theor. Phys.} {\bfseries 10} (1974) 363--384}.

\bibitem{Sushkov:2009hk}
S.~V. Sushkov, \emph{{Exact cosmological solutions with nonminimal derivative coupling}}, \href{https://doi.org/10.1103/PhysRevD.80.103505}{\emph{Phys. Rev. D} {\bfseries 80} (2009) 103505}, [\href{https://arxiv.org/abs/0910.0980}{{\ttfamily 0910.0980}}].

\bibitem{Germani:2010gm}
C.~Germani and A.~Kehagias, \emph{{New Model of Inflation with Non-minimal Derivative Coupling of Standard Model Higgs Boson to Gravity}}, \href{https://doi.org/10.1103/PhysRevLett.105.011302}{\emph{Phys. Rev. Lett.} {\bfseries 105} (2010) 011302}, [\href{https://arxiv.org/abs/1003.2635}{{\ttfamily 1003.2635}}].

\bibitem{Germani:2010ux}
C.~Germani and A.~Kehagias, \emph{{Cosmological Perturbations in the New Higgs Inflation}}, \href{https://doi.org/10.1088/1475-7516/2010/05/019}{\emph{JCAP} {\bfseries 05} (2010) 019}, [\href{https://arxiv.org/abs/1003.4285}{{\ttfamily 1003.4285}}].

\bibitem{Tsujikawa:2012mk}
S.~Tsujikawa, \emph{{Observational tests of inflation with a field derivative coupling to gravity}}, \href{https://doi.org/10.1103/PhysRevD.85.083518}{\emph{Phys. Rev. D} {\bfseries 85} (2012) 083518}, [\href{https://arxiv.org/abs/1201.5926}{{\ttfamily 1201.5926}}].

\bibitem{Yang:2015pga}
N.~Yang, Q.~Fei, Q.~Gao and Y.~Gong, \emph{{Inflationary models with non-minimally derivative coupling}}, \href{https://doi.org/10.1088/0264-9381/33/20/205001}{\emph{Class. Quant. Grav.} {\bfseries 33} (2016) 205001}, [\href{https://arxiv.org/abs/1504.05839}{{\ttfamily 1504.05839}}].

\bibitem{Linde:1983gd}
A.~D. Linde, \emph{{Chaotic Inflation}}, \href{https://doi.org/10.1016/0370-2693(83)90837-7}{\emph{Phys. Lett. B} {\bfseries 129} (1983) 177--181}.

\bibitem{Gao:2014ryw}
Q.~Gao and Y.~Gong, \emph{{The challenge for single field inflation with BICEP2 result}}, \href{https://doi.org/10.1016/j.physletb.2014.05.018}{\emph{Phys. Lett. B} {\bfseries 734} (2014) 41--43}, [\href{https://arxiv.org/abs/1403.5716}{{\ttfamily 1403.5716}}].

\bibitem{Boubekeur:2005zm}
L.~Boubekeur and D.~H. Lyth, \emph{{Hilltop inflation}}, \href{https://doi.org/10.1088/1475-7516/2005/07/010}{\emph{JCAP} {\bfseries 07} (2005) 010}, [\href{https://arxiv.org/abs/hep-ph/0502047}{{\ttfamily hep-ph/0502047}}].

\bibitem{Kallosh:2022feu}
R.~Kallosh and A.~Linde, \emph{{Polynomial {\ensuremath{\alpha}}-attractors}}, \href{https://doi.org/10.1088/1475-7516/2022/04/017}{\emph{JCAP} {\bfseries 04} (2022) 017}, [\href{https://arxiv.org/abs/2202.06492}{{\ttfamily 2202.06492}}].

\bibitem{Kallosh:2019jnl}
R.~Kallosh and A.~Linde, \emph{{On hilltop and brane inflation after Planck}}, \href{https://doi.org/10.1088/1475-7516/2019/09/030}{\emph{JCAP} {\bfseries 09} (2019) 030}, [\href{https://arxiv.org/abs/1906.02156}{{\ttfamily 1906.02156}}].

\end{thebibliography}

\providecommand{\href}[2]{#2}\begingroup\raggedright\endgroup

\end{document}